\documentclass[twocolumn,amsmath,amssymb,prl,superscriptaddress]{revtex4-2}

\usepackage{graphicx}
\usepackage{color}
\usepackage{tabularx}
\usepackage{multirow}
\usepackage{booktabs}
\usepackage{gensymb}
\usepackage[T1]{fontenc}

\begin{document}

\title{Unveiling the Two-Proton Halo Character of $^{17}$Ne:\\Exclusive Measurement of Quasi-free Proton-Knockout Reactions}

\author{C.~Lehr}
\affiliation{Technische Universit\"{a}t Darmstadt,  Department of Physics, D--64289 Darmstadt, Germany}

\author{F.~Wamers}
\affiliation{Technische Universit\"{a}t Darmstadt,  Department of Physics, D--64289 Darmstadt, Germany}
\affiliation{GSI Helmholtzzentrum f\"{u}r Schwerionenforschung GmbH, D--64291 Darmstadt, Germany}

\author{F.~Aksouh}
\thanks{\emph{Present address:} Department of Physics and Astronomy, College of Science, King Saud University, P.O. Box 2455, 11451 Riyadh, KSA}
\affiliation{GSI Helmholtzzentrum f\"{u}r Schwerionenforschung GmbH, D--64291 Darmstadt, Germany}

\author{Yu.~Aksyutina}
\affiliation{GSI Helmholtzzentrum f\"{u}r Schwerionenforschung GmbH, D--64291 Darmstadt, Germany}

\author{H.~\'{A}lvarez-Pol}
\affiliation{Instituto Galego de F\'{i}sica de Altas Enerxias, Universidade de Santiago de Compostela, ES--15782 Santiago de Compostela, Spain}

\author{L.~Atar}
\affiliation{Technische Universit\"{a}t Darmstadt,  Department of Physics, D--64289 Darmstadt, Germany}
\affiliation{GSI Helmholtzzentrum f\"{u}r Schwerionenforschung GmbH, D--64291 Darmstadt, Germany}

\author{T.~Aumann}
\email[\emph{E-mail:}~]{taumann@ikp.tu-darmstadt.de}
\affiliation{Technische Universit\"{a}t Darmstadt,  Department of Physics, D--64289 Darmstadt, Germany}
\affiliation{GSI Helmholtzzentrum f\"{u}r Schwerionenforschung GmbH, D--64291 Darmstadt, Germany}
\affiliation{Helmholtz Research Academy for FAIR, D--64289 Darmstadt, Germany}

\author{S.~Beceiro-Novo}
\thanks{\emph{Present address:} NSCL, Michigan State University, East Lansing, Michigan 48824, USA}
\affiliation{Instituto Galego de F\'{i}sica de Altas Enerxias, Universidade de Santiago de Compostela, ES--15782 Santiago de Compostela, Spain}

\author{C.A.~Bertulani}
\affiliation{Texas A\&M University-Commerce, Commerce, USA}

\author{K.~Boretzky}
\affiliation{GSI Helmholtzzentrum f\"{u}r Schwerionenforschung GmbH, D--64291 Darmstadt, Germany}

\author{M.J.G.~Borge} 
\affiliation{Instituto de Estructura de la Materia, CSIC, ES--28006 Madrid, Spain}

\author{C.~Caesar}
\affiliation{Technische Universit\"{a}t Darmstadt,  Department of Physics, D--64289 Darmstadt, Germany}
\affiliation{GSI Helmholtzzentrum f\"{u}r Schwerionenforschung GmbH, D--64291 Darmstadt, Germany}

\author{M.~Chartier}
\affiliation{Department of Physics, University of Liverpool, Liverpool L69 3BX, United Kingdom}

\author{A.~Chatillon}
\affiliation{GSI Helmholtzzentrum f\"{u}r Schwerionenforschung GmbH, D--64291 Darmstadt, Germany}

\author{L.V.~Chulkov}
\affiliation{GSI Helmholtzzentrum f\"{u}r Schwerionenforschung GmbH, D--64291 Darmstadt, Germany}
\affiliation{NRC Kurchatov Institute, RU--123182 Moscow, Russia}

\author{D.~Cortina-Gil}
\affiliation{Instituto Galego de F\'{i}sica de Altas Enerxias, Universidade de Santiago de Compostela, ES--15782 Santiago de Compostela, Spain}

\author{P.~D\'iaz Fern\'andez}
\affiliation{Instituto Galego de F\'{i}sica de Altas Enerxias, Universidade de Santiago de Compostela, ES--15782 Santiago de Compostela, Spain}
\affiliation{Department of Physics, Chalmers Tekniska H{\"o}gskola, SE--41296 G{\"o}teborg, Sweden}

\author{H.~Emling}
\affiliation{GSI Helmholtzzentrum f\"{u}r Schwerionenforschung GmbH, D--64291 Darmstadt, Germany}

\author{O.~Ershova}
\affiliation{GSI Helmholtzzentrum f\"{u}r Schwerionenforschung GmbH, D--64291 Darmstadt, Germany}
\affiliation{Goethe Universit\"{a}t Frankfurt, Department of Physics, D--60438 Frankfurt am Main, Germany}

\author{L.M.~Fraile}
\affiliation{Department of Atomic, Molecular and Nuclear Physics, Universidad Complutense de Madrid, ES--28040 Marid, Spain}

\author{H.O.U.~Fynbo}
\affiliation{Department of Physics and Astronomy, University of Aarhus, DK--8000 Aarhus, Denmark}

\author{D.~Galaviz}
\affiliation{Instituto de Estructura de la Materia, CSIC, ES--28006 Madrid, Spain}

\author{H.~Geissel}
\affiliation{GSI Helmholtzzentrum f\"{u}r Schwerionenforschung GmbH, D--64291 Darmstadt, Germany}

\author{M.~Heil}
\affiliation{GSI Helmholtzzentrum f\"{u}r Schwerionenforschung GmbH, D--64291 Darmstadt, Germany}

\author{M.~Heine}
\affiliation{Technische Universit\"{a}t Darmstadt,  Department of Physics, D--64289 Darmstadt, Germany}

\author{D.H.H.~Hoffmann}
\affiliation{Technische Universit\"{a}t Darmstadt,  Department of Physics, D--64289 Darmstadt, Germany}

\author{M.~Holl}
\affiliation{Technische Universit\"{a}t Darmstadt,  Department of Physics, D--64289 Darmstadt, Germany}
\affiliation{GSI Helmholtzzentrum f\"{u}r Schwerionenforschung GmbH, D--64291 Darmstadt, Germany}
\affiliation{Department of Physics, Chalmers Tekniska H{\"o}gskola, SE--41296 G{\"o}teborg, Sweden}

\author{H.T.~Johansson}
\affiliation{Department of Physics, Chalmers Tekniska H{\"o}gskola, SE--41296 G{\"o}teborg, Sweden}

\author{B.~Jonson}
\affiliation{Department of Physics, Chalmers Tekniska H{\"o}gskola, SE--41296 G{\"o}teborg, Sweden}

\author{C.~Karagiannis}
\affiliation{GSI Helmholtzzentrum f\"{u}r Schwerionenforschung GmbH, D--64291 Darmstadt, Germany}

\author{O.A.~Kiselev}
\affiliation{GSI Helmholtzzentrum f\"{u}r Schwerionenforschung GmbH, D--64291 Darmstadt, Germany}

\author{J.V.~Kratz}
\affiliation{Institut f\"{u}r Kernchemie, Johannes Gutenberg-Universit{\"a}t Mainz, D--55122 Mainz, Germany}

\author{R.~Kulessa}
\affiliation{Instytut Fizyki, Uniwersytet Jagello{\'n}ski, PL--30-059 Krak{\'o}v, Poland}

\author{N.~Kurz}
\affiliation{GSI Helmholtzzentrum f\"{u}r Schwerionenforschung GmbH, D--64291 Darmstadt, Germany}

\author{C.~Langer}
\thanks{\emph{Present address:} FH Aachen University of Applied Science, D--52066 Aachen, Germany}
\affiliation{GSI Helmholtzzentrum f\"{u}r Schwerionenforschung GmbH, D--64291 Darmstadt, Germany}
\affiliation{Goethe Universit\"{a}t Frankfurt, Department of Physics, D--60438 Frankfurt am Main, Germany}

\author{M.~Lantz}
\thanks{\emph{Present address:} Department of Physics and Astronomy, Uppsala University, SE--751 20 Uppsala, Sweden}
\affiliation{Department of Physics, Chalmers Tekniska H{\"o}gskola, SE--41296 G{\"o}teborg, Sweden}

\author{T.~Le~Bleis}
\affiliation{GSI Helmholtzzentrum f\"{u}r Schwerionenforschung GmbH, D--64291 Darmstadt, Germany}

\author{R.~Lemmon}
\affiliation{ Nuclear Physics Group, STFC Daresbury Lab, Warrington WA4 4AD, Cheshire, UK}

\author{Yu.A.~Litvinov}
\affiliation{GSI Helmholtzzentrum f\"{u}r Schwerionenforschung GmbH, D--64291 Darmstadt, Germany}

\author{B.~L\"{o}her}
\affiliation{GSI Helmholtzzentrum f\"{u}r Schwerionenforschung GmbH, D--64291 Darmstadt, Germany}
\affiliation{Technische Universit\"{a}t Darmstadt,  Department of Physics, D--64289 Darmstadt, Germany}

\author{K.~Mahata}
\affiliation{GSI Helmholtzzentrum f\"{u}r Schwerionenforschung GmbH, D--64291 Darmstadt, Germany}
\affiliation{Nuclear Physics Division, Bhabha Atomic Research Centre, Trombay, Mumbai-400 085, India}

\author{J.~Marganiec-Gal\k{a}zka}
\thanks{\emph{Present address:} National Centre for Nuclear Research, Radioisotope Centre POLATOM, 05-400 Otwock, Poland}
\affiliation{Technische Universit\"{a}t Darmstadt,  Department of Physics, D--64289 Darmstadt, Germany}
\affiliation{GSI Helmholtzzentrum f\"{u}r Schwerionenforschung GmbH, D--64291 Darmstadt, Germany}

\author{C.~M{\"u}ntz}
\affiliation{Goethe Universit\"{a}t Frankfurt, Department of Physics, D--60438 Frankfurt am Main, Germany}

\author{T.~Nilsson}
\affiliation{Department of Physics, Chalmers Tekniska H{\"o}gskola, SE--41296 G{\"o}teborg, Sweden}

\author{C.~Nociforo}
\affiliation{GSI Helmholtzzentrum f\"{u}r Schwerionenforschung GmbH, D--64291 Darmstadt, Germany}

\author{W.~Ott}
\thanks{deceased}
\affiliation{GSI Helmholtzzentrum f\"{u}r Schwerionenforschung GmbH, D--64291 Darmstadt, Germany}

\author{V.~Panin}
\affiliation{GSI Helmholtzzentrum f\"{u}r Schwerionenforschung GmbH, D--64291 Darmstadt, Germany}
\affiliation{Technische Universit\"{a}t Darmstadt,  Department of Physics, D--64289 Darmstadt, Germany}

\author{S.~Paschalis}
\thanks{\emph{Present address:} Department of Physics, University of York, York, YO10 5DD, UK}
\affiliation{GSI Helmholtzzentrum f\"{u}r Schwerionenforschung GmbH, D--64291 Darmstadt, Germany}
\affiliation{Department of Physics, University of Liverpool, Liverpool L69 3BX, United Kingdom}

\author{A.~Perea}
\affiliation{Instituto de Estructura de la Materia, CSIC, ES--28006 Madrid, Spain}

\author{R.~Plag}
\affiliation{GSI Helmholtzzentrum f\"{u}r Schwerionenforschung GmbH, D--64291 Darmstadt, Germany}
\affiliation{Goethe Universit\"{a}t Frankfurt, Department of Physics, D--60438 Frankfurt am Main, Germany}

\author{R.~Reifarth}
\affiliation{Goethe Universit\"{a}t Frankfurt, Department of Physics, D--60438 Frankfurt am Main, Germany}
\affiliation{GSI Helmholtzzentrum f\"{u}r Schwerionenforschung GmbH, D--64291 Darmstadt, Germany}

\author{A.~Richter}
\affiliation{Technische Universit\"{a}t Darmstadt,  Department of Physics, D--64289 Darmstadt, Germany}

\author{K.~Riisager}
\affiliation{Department of Physics and Astronomy, University of Aarhus, DK--8000 Aarhus, Denmark}

\author{C.~Rodriguez-Tajes}
\affiliation{Instituto Galego de F\'{i}sica de Altas Enerxias, Universidade de Santiago de Compostela, ES--15782 Santiago de Compostela, Spain}

\author{D.~Rossi}
\affiliation{Technische Universit\"{a}t Darmstadt,  Department of Physics, D--64289 Darmstadt, Germany}
\affiliation{GSI Helmholtzzentrum f\"{u}r Schwerionenforschung GmbH, D--64291 Darmstadt, Germany}
\affiliation{Institut f\"{u}r Kernchemie, Johannes Gutenberg-Universit{\"a}t Mainz, D--55122 Mainz, Germany}

\author{D.~Savran}
\affiliation{GSI Helmholtzzentrum f\"{u}r Schwerionenforschung GmbH, D--64291 Darmstadt, Germany}

\author{H.~Scheit}
\affiliation{Technische Universit\"{a}t Darmstadt,  Department of Physics, D--64289 Darmstadt, Germany}

\author{G.~Schrieder}
\affiliation{Technische Universit\"{a}t Darmstadt,  Department of Physics, D--64289 Darmstadt, Germany}

\author{P.~Schrock}
\affiliation{Technische Universit\"{a}t Darmstadt,  Department of Physics, D--64289 Darmstadt, Germany}

\author{H.~Simon}
\affiliation{GSI Helmholtzzentrum f\"{u}r Schwerionenforschung GmbH, D--64291 Darmstadt, Germany}

\author{J.~Stroth}
\affiliation{Goethe Universit\"{a}t Frankfurt, Department of Physics, D--60438 Frankfurt am Main, Germany}
\affiliation{GSI Helmholtzzentrum f\"{u}r Schwerionenforschung GmbH, D--64291 Darmstadt, Germany}

\author{K.~S{\"u}mmerer}
\affiliation{GSI Helmholtzzentrum f\"{u}r Schwerionenforschung GmbH, D--64291 Darmstadt, Germany}

\author{O.~Tengblad}
\affiliation{Instituto de Estructura de la Materia, CSIC, ES--28006 Madrid, Spain}

\author{H.~Weick}
\affiliation{GSI Helmholtzzentrum f\"{u}r Schwerionenforschung GmbH, D--64291 Darmstadt, Germany}

\author{C.~Wimmer}
\affiliation{Goethe Universit\"{a}t Frankfurt, Department of Physics, D--60438 Frankfurt am Main, Germany}
\affiliation{GSI Helmholtzzentrum f\"{u}r Schwerionenforschung GmbH, D--64291 Darmstadt, Germany}

\date{\today}

\begin{abstract}
The proton drip-line nucleus $^{17}$Ne is investigated experimentally in order to determine its two-proton halo character. A fully exclusive measurement of the $^{17}$Ne$(p,2p)^{16}$F$^*\rightarrow ^{15}$O$+p$ quasi-free one-proton knockout reaction has been performed at GSI at around 500~MeV/nucleon beam energy. All particles resulting from the scattering process have been detected. The relevant reconstructed quantities are the angles of the two protons scattered in qusi-elastic kinematics, the decay of $^{16}$F into $^{15}$O (including $\gamma$ decays from excited states) and a proton, as well as the  $^{15}$O$+p$ relative-energy spectrum and the $^{16}$F momentum distributions. The latter two quantities allow an independent and consistent determination of the ratio of $l=0$ and $l=2$ motion of the valence protons in $^{17}$Ne. With a resulting relatively small $l=0$ component of only around 35(3)\%, it is concluded that $^{17}$Ne exhibits a rather modest halo character only. The quantitative agreement of the two values deduced from the energy spectrum and the momentum distributions supports the theoretical treatment of the calculation of momentum distributions after quasi-free knockout reactions at high energies by taking into account distortions based on the Glauber theory. Moreover, the experimental data allow the separation of valence-proton knockout and knockout from the $^{15}$O core. The latter process contributes with 11.8(3.1)~mb around 40\% to the total proton-knockout cross section of 30.3(2.3)~mb, which explains previously reported contradicting conclusions derived from inclusive cross sections.
\end{abstract}


\maketitle

{\it Introduction}.---Atomic nuclei at the limits of nuclear binding, located close to the neutron and proton drip lines, exhibit unusual and often unexpected properties when compared to expectations from theoretical models, which describe known properties of ordinary stable or very long-lived nuclei well. Such exotic nuclear properties can originate from a strong imbalance in the neutron-to-proton ratio amplifying subtle properties of the nuclear interaction, which are not decisive for the understanding of stable nuclei. Because of the closeness of the continuum, such nuclei are called `open quantum systems', often dominated by correlations~\cite{ Michel2008}. The experimental study of properties of drip-line nuclei is thus key for developing and testing modern nuclear theory and the interactions used. The knowledge of properties of exotic nuclei, either reliably theoretically predicted or directly measured, is also a basis for the understanding of nucleosynthesis processes in astrophysics, such as the rapid proton- and neutron-capture processes.      

Particular attention in this direction was dedicated to halo nuclei since their experimental discovery at the Berkeley BEVALAC by Tanihata {\it et al.}~\cite{Tanihata1985a,Tanihata1985b}  and the following interpretation and name coining nuclear `halo' by Hansen and Jonson~\cite{HanJon1987}. Halo nuclei exhibit a matter-density distribution with a pronounced, far-extending, low-density tail (halo), which is caused by the wavefunction of the last weakly-bound valence nucleons, which reaches far into the classical forbidden region. This generates an almost pure neutron or proton low-density environment at the surface of the nucleus. On the neutron drip-line, several such nuclei have been studied in great detail, thanks to the tremendous progress in rare-isotope beam accelerator facilities and experimental instrumentation. A prime example is $^{11}$Li with two loosely-bound neutrons forming the halo, where the two neutrons are strongly correlated in a mixed ground-state wavefunction comprising components with angular momenta $l=0$ and $l=1$~\cite{Simon1999, Kubota2020}. A sizable low-angular-momentum component is essential for the formation of a halo-like structure. 

On the much-less studied proton drip-line side, $^{17}$Ne is the most promising candidate for such a structure. In a three-body model,  $^{17}$Ne can be described as a well bound $^{15}$O core plus two protons loosely bound with a two-proton separation energy of only $S_{2p}=933.1(6)$~keV~\cite{Wang2012}. According to the standard nuclear shell model, the oxygen core has a closed proton shell, with the next available states in the $sd$ shell. This suggests the possibility of halo formation with the two protons in $l=0$ motion~\cite{Zhukov1995}. Moreover, like $^{11}$Li, with both sub-systems ($^{16}$F and $p-p$)  being unbound, $^{17}$Ne is a \emph{Borromean} nucleus.
Albeit numerous experimental efforts, a firm conclusion on the structure of  $^{17}$Ne has not yet been reached. In this Letter, we present experimental results from an exclusive measurement of the $(p,2p)$ proton knockout reaction. The ratio of the $s$ and $d$ components in the valence-nucleon motion has been determined by two independently measured quantities, providing a clear answer to this long-standing discussion. With a resulting $s^2$ to $d^2$ ratio of around $\frac{1}{3}$ to $\frac{2}{3}$, it is concluded that $^{17}$Ne exhibits only a rather moderate halo character.

{\it Summary of the $^{17}\mathrm{Ne}$ puzzle}.---$^{17}\mathrm{Ne}$ ($J^{\pi}$ = $1/2^{-}$) has seven neutrons and ten protons, decays ($\beta^+$) towards $^{17}$F ($T_{1/2}$ = 109~ms), and is loosely bound with a two-proton separation energy of $S_{2p}=933.1(6)$~keV, while its neutron separation energy is $S_n=15557(20)$~keV~\cite{Wang2012}. 
Evidence for a proton halo in $^{17}\mathrm{Ne}$, {\it i.e.},  a dominance of the $(s_{1/2})^2$ configuration was claimed from measurements of total interaction cross sections for $A=17$ high-energy beams~\cite{Ozawa1994}. However, calculations of the interaction cross section based on Hartree-Fock-type wave functions and the Glauber model result in the opposite conclusion with a dominance of the $(1d_{5/2})^2$ configuration~\cite{Kitagawa1997}.  
An ($l$=0)-dominating two-proton halo structure of $^{17}\mathrm{Ne}$ was also inferred from an analysis of the $^{15}\mathrm{O}$ momentum distribution and cross section for the inclusive $2p$-removal reaction at 66~MeV/nucleon~\cite{Kanungo2003,Kanungo2004}, while a shell-model interpretation of the magnetic-moment measurement~\cite{Geithner2005} arrives at the opposite conclusion. Furthermore, the measured charge radii for $^{17,18,19}$Ne of  3.04(2), 2.97(2), and 3.01(1)~fm~\cite{Geithner2008} do not support a pronounced halo for  $^{17}$Ne.

Results from theoretical predictions span the full range from $s^2$ dominance to $d^2$ dominance. Calculations within the framework of a three-cluster generator-coordinate model resulted in a dominant $(1s_{1/2})^2$ configuration~\cite{Timofeyuk1996}, while calculations using a density-dependent contact pairing interaction predicted values of $P((1s_{1/2})^2)=15.2$\%, $P((0d_{5/2})^2)=75.2$\%, and $P((0d_{3/2})^2)=3.8$\%~\cite{Oishi2010,Oishi2010_E}. A $^{15}\mathrm{O}+p+p$ three-body model was used to calculate Thomas-Ehrman shifts for $^{17}\mathrm{Ne}$ and $^{17}\mathrm{N}$~\cite{Grigorenko2003,Grigorenko2004,Garrido2004_1} resulting in $P((1s_{1/2})^2)$ values of 40-50\%. As mentioned in Ref.~\cite{Garrido2004_1}, the computed three-body Thomas-Ehrman shifts are relatively inaccurate. Coulomb energies for mirror nuclei $^{17}\mathrm{Ne}$ and $^{17}\mathrm{N}$ were computed in other models~\cite{Nakamura1998,Fortune2001,Fortune2006}. While the predominance of a $(1s_{1/2})^2$ configuration was stated in Ref.~\cite{Nakamura1998}, the two other publications agree on $P((1s_{1/2})^2)=24(3)$\%. 

Clearly, a more exclusive measurement of observables with high sensitivity and selectivity to distinguish the  $l=0$ and $l=2$  contributions for the two weakly-bound valence protons is mandatory to conclude on the halo character of  $^{17}\mathrm{Ne}$. The quasi-free proton-knockout reaction described in this Letter provides this sensitivity. The populated resonances with known structure are identified via invariant-mass spectroscopy of the residual fragment. The shape of the measured momentum distribution is sensitive to the slope of the nuclear density distribution at the surface, which is dominated by the exponential decay of the least-bound nucleon's wavefunction, which strongly depends on the angular momentum due to the angular-momentum barrier. The different shapes of the $s$ and $d$ density distributions at the surface result in different shapes of the measured momentum distributions. The extraction of a $s$ to $(s+d)$ ratio is thus rather direct and less model dependent.

{\it Experiment}.---The primary $^{20}$Ne beam was extracted from the GSI synchrotron SIS18 with an energy of 630~MeV/nucleon and directed to the production target at the fragment separator FRS. The secondary $^{17}$Ne beam entered the experimental area Cave C at GSI with an average energy of 498~MeV/nucleon in the middle of the CH$_2$ target. The average intensity of the secondary beam amounted to $10^4$/s with a $^{17}$Ne content of more than 90\% before selection. 
The beam energy of 500~MeV/nucleon was chosen to minimize secondary reactions of protons in the nuclear medium after the primary $pp$ scattering process. The energy of outgoing protons is in average 250 MeV at scattering angles of around $45\degree$, for which the nucleon-nucleon ($NN$) cross section is minimal. 
\begin{figure}
\includegraphics[width=\columnwidth]{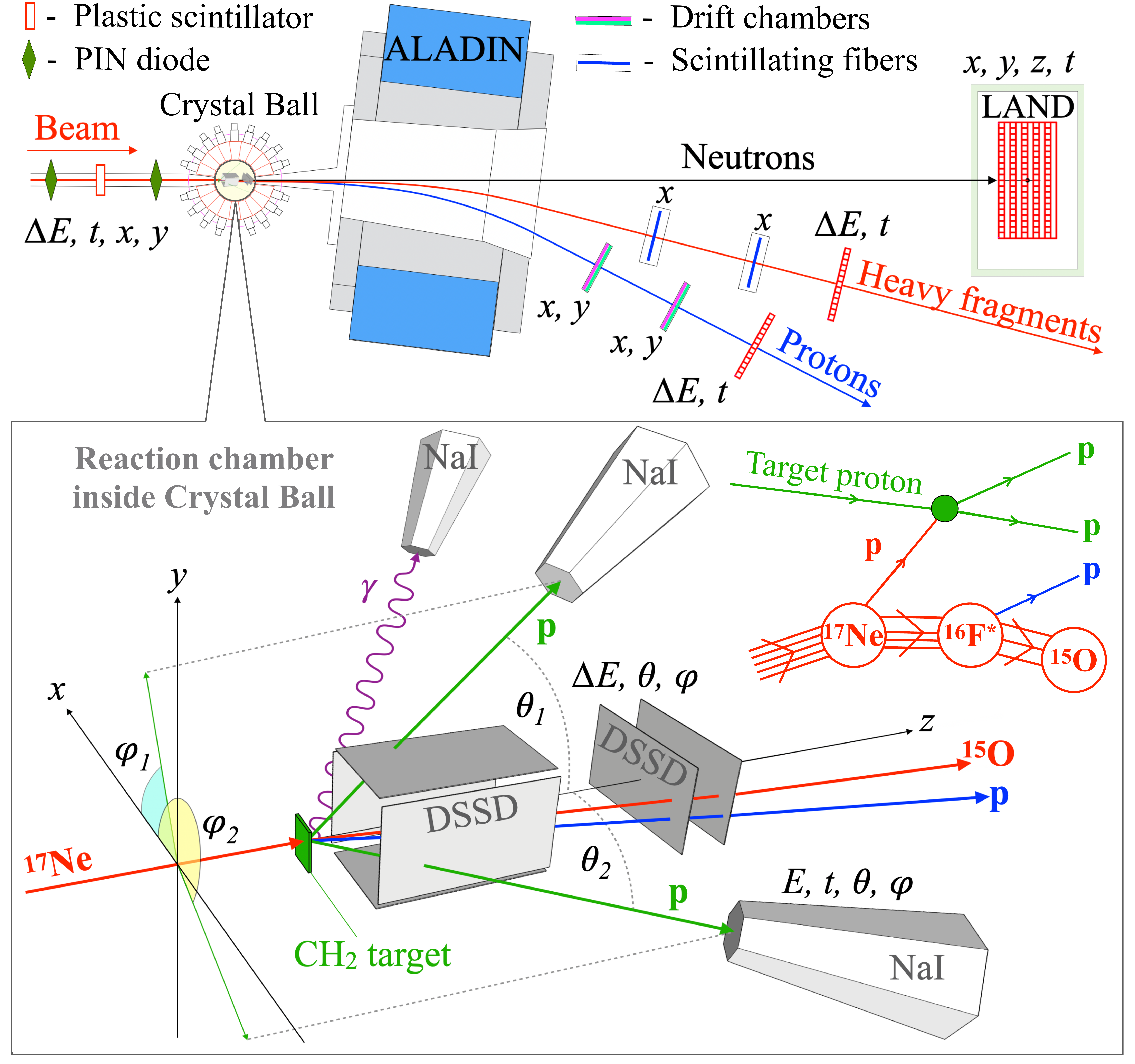}
\caption{Schematic drawing of the experimental setup (not to scale). The upper part indicates the detection systems and measured quantities to track and identify projectiles and forward-emitted reaction products. The lower frame provides a more detailed view of the detection systems around the target and the reaction studied. Photons are detected with NaI Crystals, protons with double-sided Si-strip detectors (DSSD) and NaI crystals.}
\label{setup}
\end{figure}
The experimental setup is schematically shown in Fig.~\ref{setup}, and is identical to the one described in Refs.~\cite{Wamers2014,Panin2016,Wamers2018}. The quasi-free one-proton knockout reaction $^{17}$Ne$(p,2p)^{16}$F$^*\rightarrow ^{15}$O$+p$ has been analyzed. Measurements with 213~mg/cm$^2$ CH$_2$ and 370~mg/cm$^2$ C targets have been performed as well as a measurement without target in order to determine background from reactions in other materials outside the target.

Identified $^{17}$Ne incoming ions are tracked with position-sensitive silicon PIN diodes towards the target. After the reaction target, outgoing fragments are deflected in the large-gap dipole magnet ALADIN and identified. Their angles, velocity, and magnetic rigidity are determined by double-sided silicon micro-strip detectors (DSSD), scintillating fibre detectors, and a time-of-flight (ToF) wall.  For the reaction channel of interest, identified $^{15}$O fragments have been selected. 

\begin{figure}
\includegraphics[width=\columnwidth]{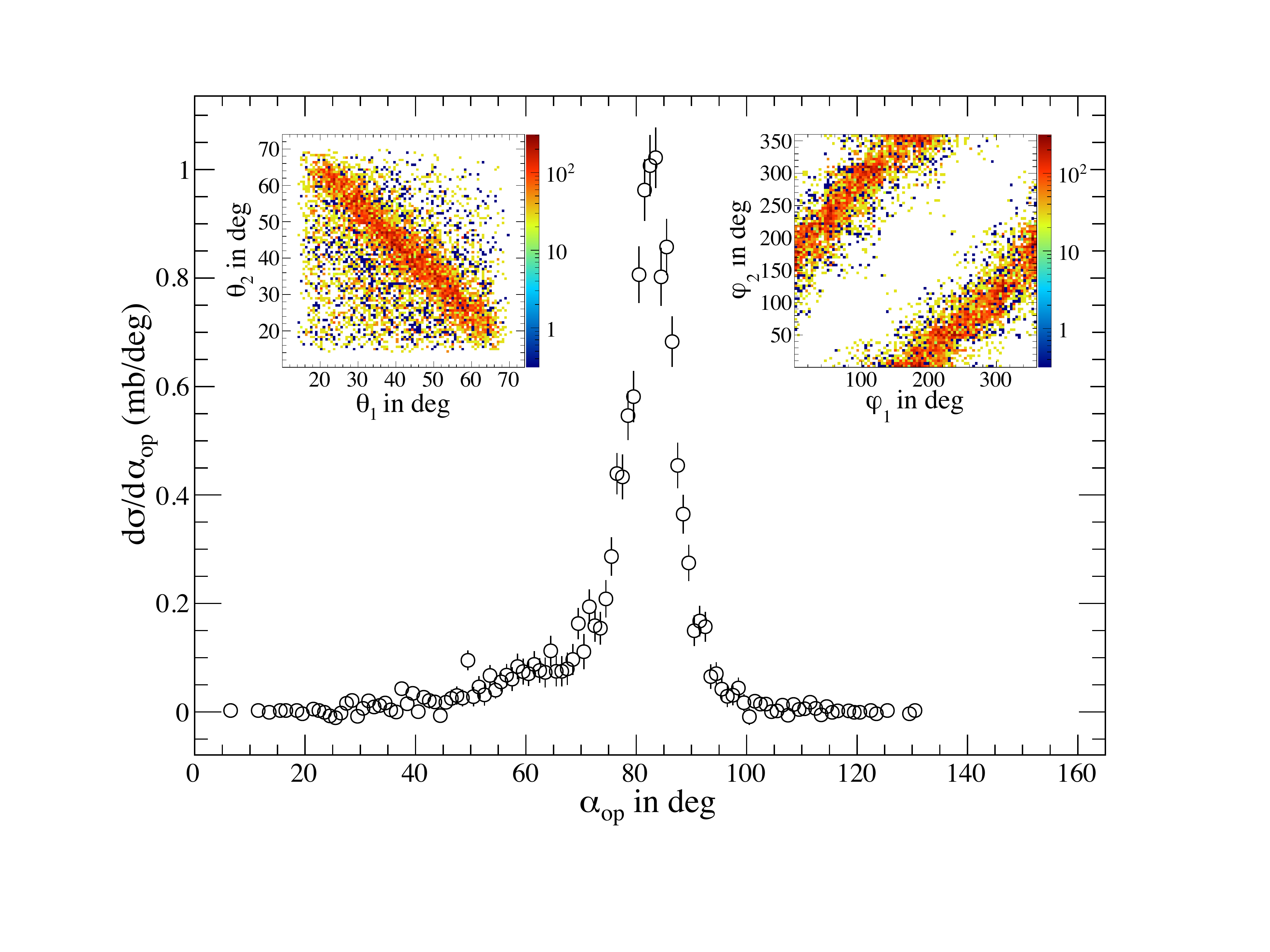}
\caption{Angular correlations of the two scattered protons in the $(p,2p)$ knockout reaction. The insets display the correlations between the polar (left) and azimuthal (right) angles. The main figure shows the distribution as function of the projected opening angle $\alpha_{op}=\theta_1+\theta_2$ of the two protons.
}
\label{pp-angles}
\end{figure}

The $(p,2p)$ reaction channel is further characterized by the measurement of the angular distribution of the two scattered protons (including the target proton), and the forward emitted proton from the decay of unbound $^{16}$F states populated after proton knockout. The scattered protons are detected by a box of four DSSDs covering an angular range of 15$\degree$ to 72$\degree$, and the Crystal Ball (CB) consisting of 162 individual NaI crystals surrounding the target (see lower part of Fig.~\ref{setup}). The resulting angular distributions of scattered protons are displayed in Fig.~\ref{pp-angles} (see lower part of Fig.~\ref{setup} for the definition of angles). They exhibit a back-to-back scattering with an opening angle peaking at around  $84\degree$ as expected for quasi-free $NN$ scattering. Deviations from elastic $pp$ scattering like the width of the distributions and the slightly reduced average opening angle have their origin in the internal motion of the proton in $^{17}$Ne and its binding energy relative to that of the the final state. 

Protons originating from the decay in flight of $^{16}$F are emitted in the laboratory frame at forward angles and tracked through the dipole magnet by two DSSDs located before the magnet, and drift chambers plus a ToF wall after the magnet. $\gamma$ decays from excited states are detected by the CB spectrometer and calorimeter.

\begin{figure}[t]
\includegraphics[width=\columnwidth]{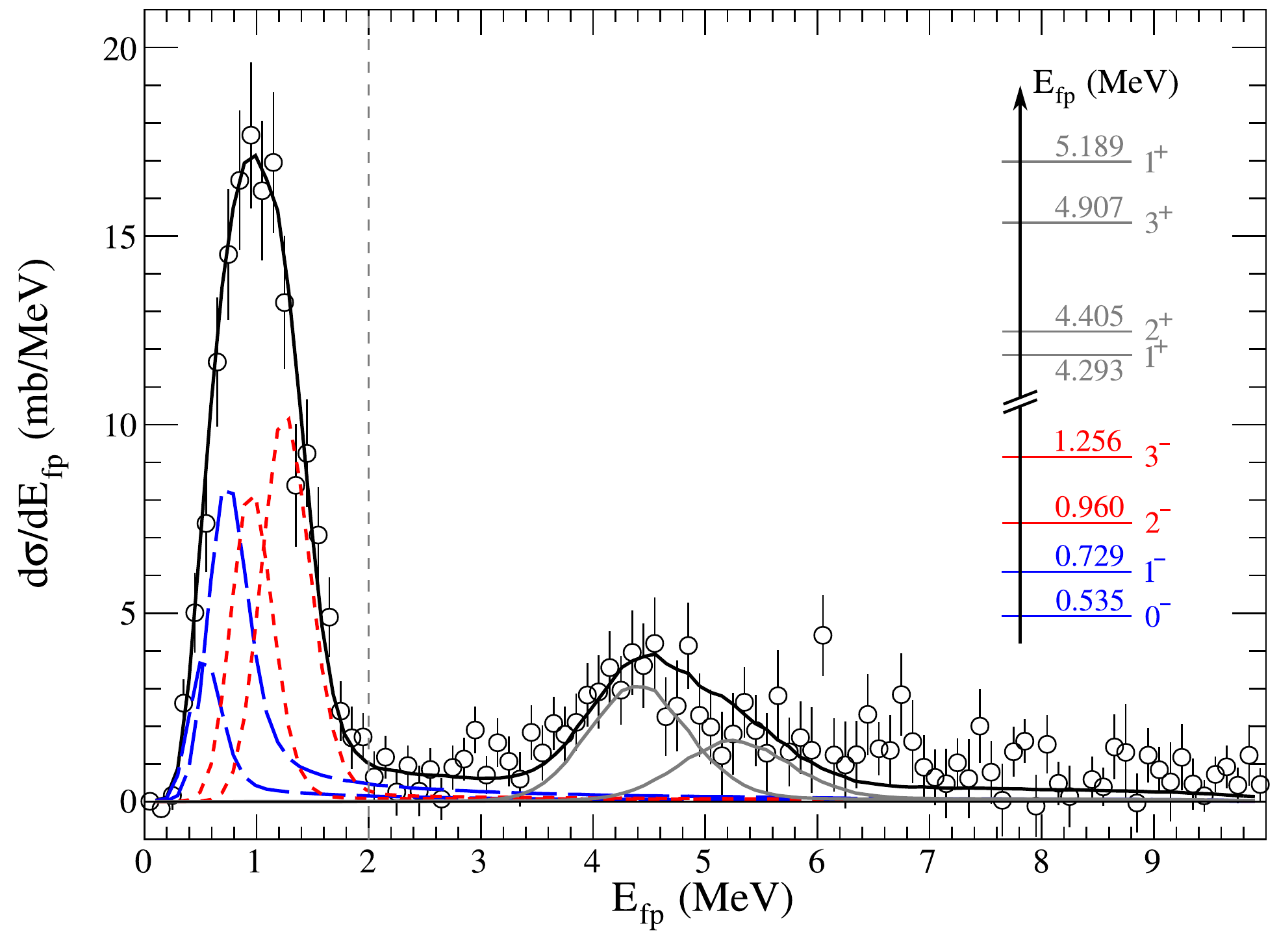}
\caption{Differential cross section $d\sigma/dE_\text{fp}$ as function of the relative energy $E_\text{fp}$ between the $^{15}$O fragment and the decay proton for the reaction $^{17}$Ne$(p,2p)^{16}$F$^*\rightarrow ^{15}$O$+p$. Contributions involving $\gamma$-decays have been subtracted. The prominent peak below 2~MeV results from four low-lying resonances in $^{16}$F populated after knockout of a valence proton from $^{17}$Ne, while the cumulation of events between 3 and 6~MeV corresponds to knockout of a proton from the $^{15}$O core. The inset shows the populated states with their energy and quantum numbers.
}
\label{erel}
\end{figure}

{\it Results}.---The resulting differential cross section $d\sigma/dE_\text{fp}$ as function of the relative-energy $E_\text{fp}$ between the $^{15}$O fragment and the decay proton for the reaction $^{17}$Ne$(p,2p)^{16}$F$^*\rightarrow ^{15}$O$+p$ is shown in Fig.~\ref{erel}, where contributions of around 10\% associated with additional $\gamma$-decays were subtracted. The spectrum exhibits two clear structures. The peak structure at higher energies reflects the population of high-lying excited states in $^{16}$F after knockout of a proton from the $^{15}$O core in  $^{17}$Ne. Together with the decays accompanied by additional $\gamma$ decays, the core knockout reaction contributes with 11.8(3.1)~mb around 40\% to the total one-proton knockout cross section $\sigma_{(p,2p)} = 30.3(2.3)$~mb. The low-lying peak results from the population of single-particle states of $^{16}$F which are not resolved. The black solid line shows a fit to the data consisting of a sum of Breit-Wigner curves with known energies and widths of the resonances adopted from the literature~\cite{Tilley1993}. 
The two low-lying $0^-$ and $1^-$ states (see level scheme shown in Fig.~\ref{erel}) are $s$-wave resonances populated after the knockout of a valence proton from the $s^2$ configuration, while the two higher lying $2^-$ and $3^-$ states are the $d$-resonances in  $^{16}$F populated after knockout from the $d^2$ configuration. The low-energy part of the spectrum below 2~MeV corresponds thus to the valence-proton or `halo' knockout. The fit results in a cross section of 18.5(2.1)~mb for the halo knockout with a relative contribution for the $s$-states of 42(5)\% The high-energy positive-parity states of the spectrum above 2~MeV are populated in core knockout reactions. The grey curves indicates the fit result, wehre the two groups of non-resolved resonances are approximated by two Breit-Wigner curves.  

The experimental cross sections are compared to theoretical $(p,2p)$ quasi-free-scattering cross sections computed using the Glauber theory \cite{AumBert2013}. Inputs to the calculations are the $^{15}$O core density distribution, single-particle wave functions for the valence protons, and the free $NN$ cross sections.  A Hartree-Fock density is used for the core with a radius $r_{rms}=2.64$~fm. 
The single-particle wave functions were obtained by solving the Schrödinger equation for a Woods-Saxon mean-field potential with radius parameter $r_0=1.2$~fm and diffuseness $a=0.7$. Cross sections were computed individually for the angular momenta $l=0$ and $l=2$ and effective binding energies according to the resonances populated. The obtained single-particle cross sections $\sigma_{sp}$ for the $s$ and $d$ states are  11.65~mb and 9.16~mb, respectively. The cross section for $l=0$ is somewhat larger due to the surface-dominated reaction probability and the long halo-like tail of the $s$ wavefunction. This results in spectroscopic factors of $C^2S=0.67(11)$ for the $s^2$ configuration and $C^2S=1.17(16)$ for the $d^2$ configuration, where $C^2S=\sigma_{exp / }\sigma_{sp}$. The probability $P(\frac{s^2}{s^2+d^2})$ to find the two valence protons in the $s^2$ configuration in the $^{17}$Ne ground state amounts thus to 36(5)\%.  

\begin{figure}[t]
\includegraphics[width=\columnwidth]{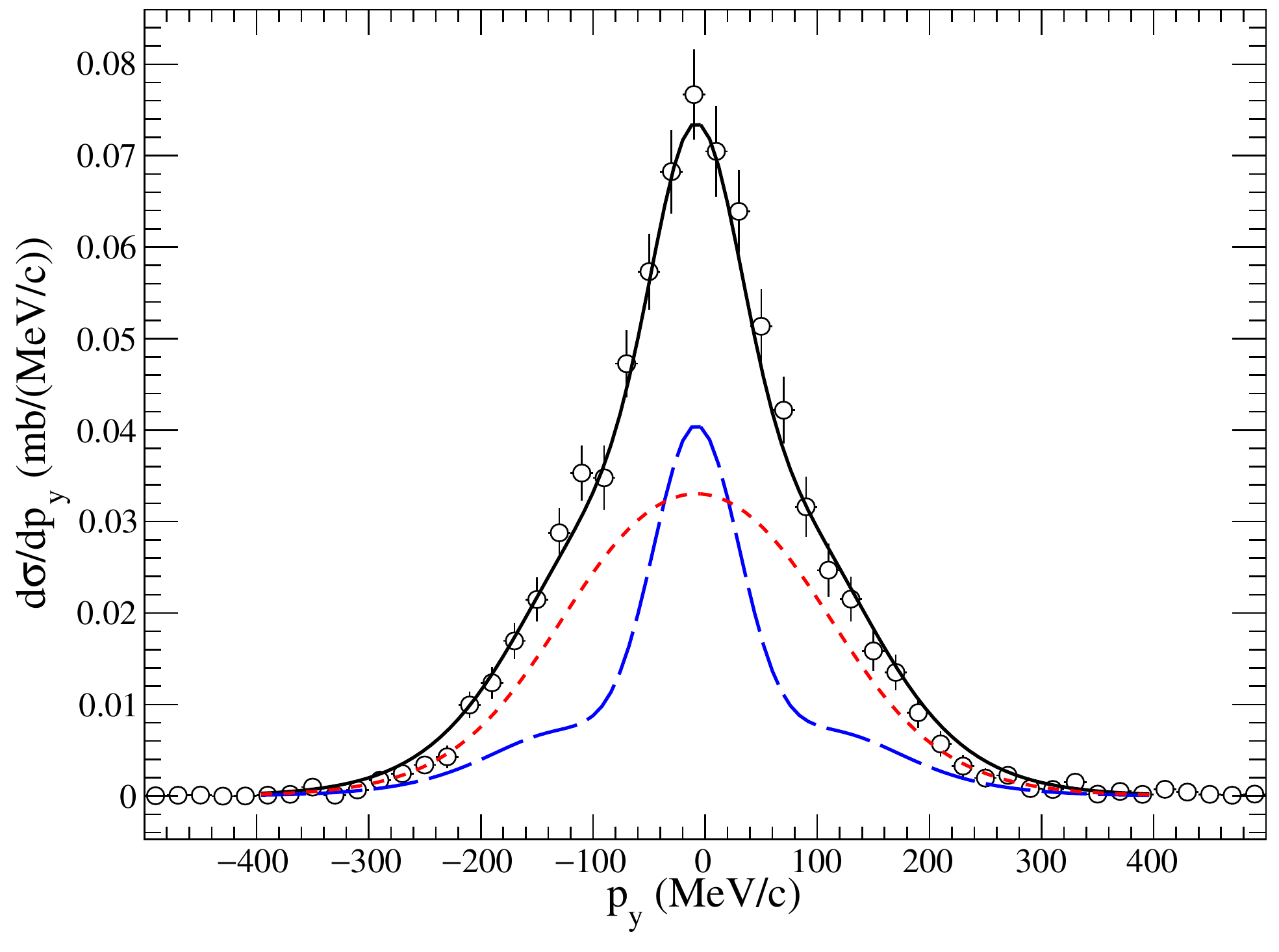}
\caption{Momentum distribution $d\sigma/dp_y$ of $^{16}$F projected onto the cartesian coordinate $y$ perpendicular to the beam for the reaction  $^{17}$Ne$(p,2p)^{16}$F$^*\rightarrow ^{15}$O$+p$ with the condition $E_\text{fp}<2$~MeV. The solid curve represents the theoretical result after adjusting the $l=0$ (long-dashed) and $l=2$ (short-dashed) contributions to the experimental data (symbols).
}
\label{py}
\end{figure}

The shape of the momentum distribution $d\sigma/dp$ of the residual fragment after one-nucleon knockout is characteristic for the angular momentum of the knocked-out nucleon. The  transverse momentum distribution $d\sigma/dp_y$ of  $^{16}$F, projected onto the cartesian coordinate $y$, is shown in Fig.~\ref{py}. The distribution was reconstructed from the measured momenta of $^{15}$O plus the forward-emitted proton from the decay of $^{16}$F, with the condition that the relative energy $E_\text{fp}< 2$~MeV, {\it i.e.}, with a selection on `halo' knockout. The data clearly indicate a superposition of two shapes. The solid curve represents a fit of the calculated distributions to the data, using the above-described theoretical description, for $l=0$ (long-dashed curve) and $l=2$ (short-dashed curve) resulting in a relative contribution to the cross section of 39(4)\% for $l=0$, corresponding to a probability of 34(3)\% for the $s^2$ configuration of the valence protons. This is in perfect agreement with the independent result derived from the relative-energy spectrum discussed above.

\begin{figure}
\includegraphics[width=\columnwidth]{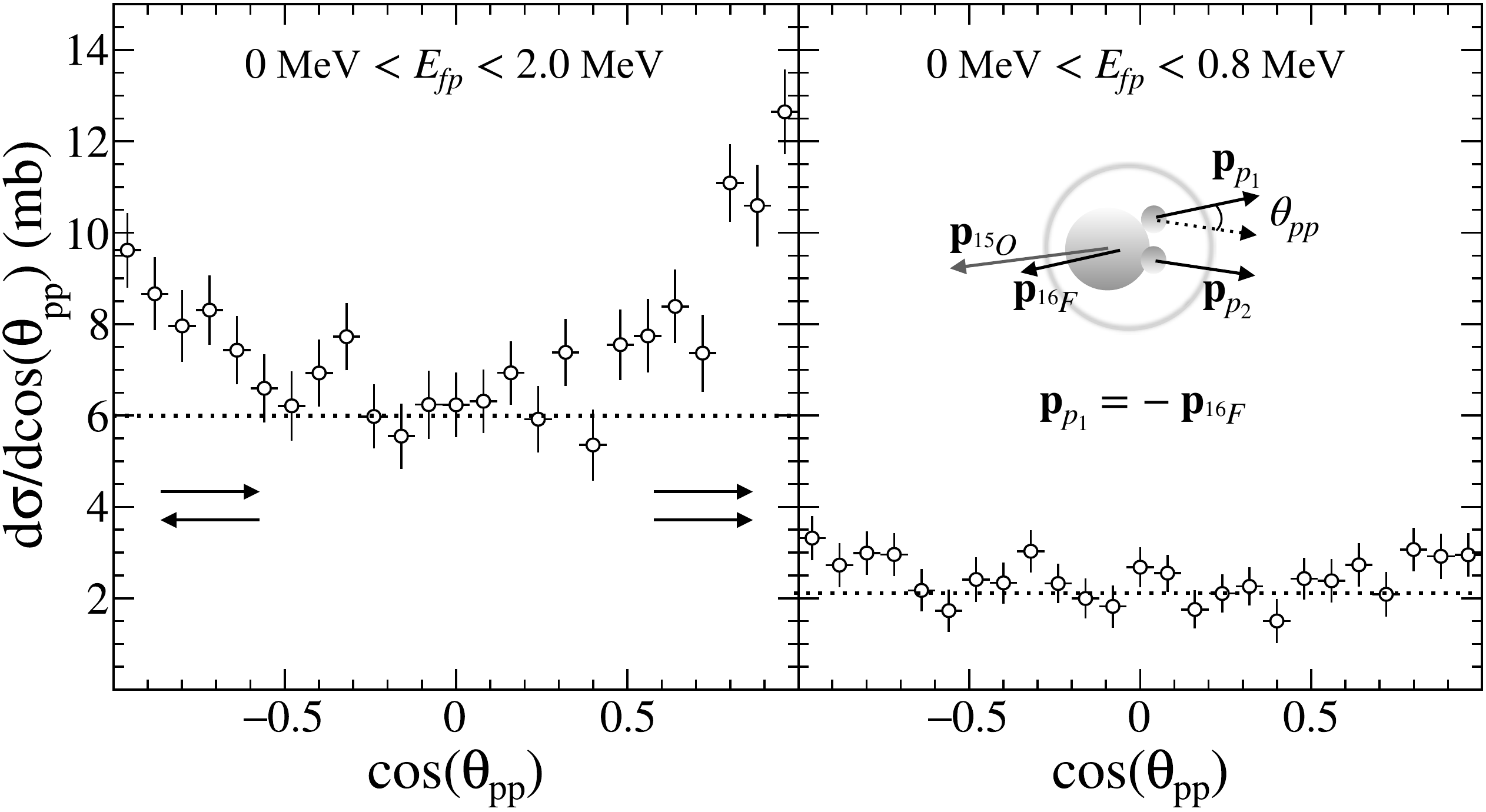}
\caption{Proton-proton angular correlation $\theta_{pp}$ of the momenta of the two halo protons in $^{17}$Ne, represented by the angle between the recoil momentum of $^{16}$F and the direction of the $^{15}$O$-p$ relative motion (see text).
The left frame shows the distribution for halo knockout with population of $^{16}$F in the energy region $E_\text{fp}<2.0\text{\,MeV}$ with overlapping $s$ and $d$ resonances, while the right frame displays the correlation for $E_\text{fp}<0.8$~MeV, where the $l=0$ configuration dominates. 
}
\label{ppcorr}
\end{figure}

The dominance of the $l=2$ configuration is further supported by the proton-proton angular correlations in the $^{17}$Ne ground state shown in  Fig.~\ref{ppcorr}. 
The $p-p$ relative angle $\theta_{pp}$ in the $^{17}$Ne frame has been constructed under the assumption that the relative motion of the fragment $^{15}$O and the proton $p_2$ (from the decay of $^{16}$F, see inset of Fig.~\ref{ppcorr}) remains undisturbed after sudden knockout of the proton $p_1$. 
The angular correlation between the protons in the $^{17}$Ne ground state is accordingly reflected by the angle between the momentum of the knocked out proton $p_1$, given by the $^{16}$F recoil with ${\bf p}_{p_1} = -{\bf p}_{^{16}F}$  (in the rest frame of the projectile), and the $^{15}$O$-p_2$ relative momentum ${\bf p}_{fp}=\mu ({\bf p}_{p_2}/m_p+{\bf p}_f/m_f)$, where $m_p$,  $m_f$, and $\mu$ are the masses of proton, $^{15}$O, and the reduced mass of the $^{16}$F system, respectively.
The distribution exhibits a strong asymmetry caused by the $d$-wave contribution with preferred back-to-back or parallel motion (Fig.~\ref{ppcorr} left frame), while for a pure $s$-state, a symmetric and isotropic distribution is expected~\cite{ChulSchr1997, Chulkov1997}. This can be seen in the right frame of Fig.~\ref{ppcorr}, where the distribution is shown for the condition $E_\text{fp}<0.8$~MeV, for which the $l=0$ contribution dominates. Only a slight asymmetry is visible caused by remaining $d$ admixtures.

{\it Conclusion}.---The structure of the proton-halo candidate $^{17}$Ne has been investigated by performing and analyzing an exclusive measurement of the $^{17}$Ne$(p,2p)^{16}$F$^*\rightarrow ^{15}$O$+p$ reaction at high beam energy of around 500~MeV/nucleon. The data allowed for the identification of quasi-free $(p,2p)$ knockout from the valence states. The analysis of two independent observables, the $^{15}$O$-p$ relative-energy spectrum and the $^{16}$F momentum distribution, results in a consistent interpretation of the structure of the $^{17}$Ne ground-state configuration, where the two valence protons occupy dominantly $s^2$ and $d^2$ configurations with a rather small $s^2$ component of 35(3)\%. The dominance of the $d^2$ contribution suppresses the halo character of $^{17}$Ne. The large total spectroscopic factor of $C^2S=1.8(2)$ indicates no or only minor contributions due to more complex configurations, and supports a description of $^{17}$Ne in a $^{15}$O$+p+p$ three-body model with an inert $^{15}$O core.

The quantitative agreement of the analysis of two independently measured quantities, the population of final states and the momentum distributions, give confidence on the accuracy of treating the $(p,2p)$ reaction in the Glauber theory based on eikonal wavefunctions as developed in Ref.~\cite{AumBert2013}. The extraction of ratios of different configurations from momentum distributions relies heavily on the calculated shape of the distributions, which is affected by distortions due to the reaction mechanism. The perfect agreement of the theoretical shape with the measured distribution in conjunction with a perfect agreement of the extracted $s^2$ ratio provides a solid basis for the investigation of exotic nuclei using quasi-free scattering at the upcoming radioactive-beam facilities FAIR and FRIB.

The controversial conclusions in the literature on the halo structure of  $^{17}$Ne can be resolved. We discuss here only the least indirect methods based on the measurement of observables exhibiting a pronounced sensitivity on the $s$-wave character of the valence-nucleon wave function: the measurement of the magnetic dipole moment~\cite{Geithner2005} and the inclusive proton-removal reaction~\cite{Kanungo2003}. The result presented in this Letter is in agreement with the shell-model interpretation of the magnetic-moment measurement. The disagreement with the interpretation of the inclusive proton-removal reaction can also be understood, since core-knockout contributions have been assumed to be small in Ref.~\cite{Kanungo2003}. The separation of the contributions of the core and valence-nucleon knockout in the exclusive experiment, however, reveals a significant contribution to the total proton-removal cross section of proton knockout from the core of around~40\%. The search for a well-developed halo nucleus at the proton dripline, comparable to neutron-halo states, still remains an open challenge.

This work is supported by the German Federal Ministry for Education and Research (BMBF) under contract No. 05P15RDFN1, the ExtreMe Matter Institute (EMMI), HIC for FAIR, and the GSI-TU Darmstadt cooperation agreement. Financial support from the Swedish Research Council, from the Russian Foundation for Basic Research (RFBR Grant 12-02-01115-a), and from the Spanish grants from MICINN AEI FPA2017-87568-P, PGC2018-0099746-B-C21,   FPA2015-646969-P, PID2019-104390GB-I00 and Maria de Maeztu Units of excellence MDM-2016-0692 are also acknowledged. One of us (B. J.) is a Helmholtz International Fellow. C.A.B. acknowledges support by the U.S. DOE grant DE-FG02-08ER41533 and the U.S. NSF Grant No. 1415656.


%

\end{document}